# Effect of Atomic Size and Valence Electron Concentration on the Formation of fcc or bcc Solid Solid Solutions in High Entropy Alloys.

O. Coreño-Alonso(1), J. Coreño-Alonso(2)

(1)Departamento de Ingeniería Civil, Universidad de Guanajuato, México.

(2) Área Académica de Materiales y Ciencias de la Tierra, UAEH, México.

**Abstract.**

The possibility of solid solution formation in high entropy alloys (HEAs) has been calculated for alloys with four to seven elements, using a rule previously reported. Thirty elements were included: transition elements of the fourth, fifth and sixth periods of the periodic table, and aluminum. A total of 2,799,486 systems were analyzed. The percentage of solid solutions that would be formed in HEAs decreases from 35.9% to 26.4%, as the number of elements increases from four to seven. The structure of the solid solutions, fcc, bcc or a mixture of fcc and bcc, that would be formed, has been predicted using a previously reported observation. The percentage of systems with fcc or bcc structure decreases as the number of elements increases from four to seven. The percentages of solid solutions with fcc, bcc or a mixture of fcc and bcc were calculated, for alloys with four to seven elements, but maintaining one constant element. Systems in which the constant element has valence electron concentration, VEC, from three to four, would have bcc structure in around 50% of the systems. Systems in which the constant element has VEC from ten to twelve, around 75% of systems would present fcc structure.

## 1. Introduction.

In conventional alloys, one principal element is combined with several alloying elements to produce alloys with different properties. For example, Fe is the base of steels, and depending upon the alloying elements, structural, stainless, tool, magnetic, wear resistant or high temperature steels can be produced. High entropy alloys (HEAs) are alloys with five or more principal elements in equal or near equal atomic percent [1,2]. HEAs can also be defined as alloys that generally have at least five major metallic elements, each of which has an atomic percentage between 5% and 35% [3].Due to the distinct design concept, these alloys often exhibit unusual properties. Thus, there has been significant interest in these materials, leading to an emerging yet exciting new field [1]. Excellent reviews exist on this field [1-2,4]. Characteristics of HEAS such as core effects, phases and crystal structures, structural stability, mechanical, wear and corrosion resistance, magnetic, chemical and physical and high temperature properties have been reported [1-11].

The relationship between structure and properties of the materials is well known. For example, the reported results show that, in general terms, fcc-structured HEAs have low strength and high plasticity. On the other hand, HEAS with a bcc structure exhibit high strength and low plasticity. This behavior is related to the fact that bcc and bcc-derivative structures are harder than their fcc counterparts because bcc based structures have stronger directional bonding and lack a truly close-packed slip plane [1]. Although, the production of a HEA with fine lamellar fcc/B2 microstructure, and which showed an unprecedented combination of high tensile ductility and high fracture strength at room temperature has been reported [12]. Then, the possibility to predict phases in HEAs could lead to the production of HEAs with desired properties. However, up to now, only a few studies have been carried out on phase formation rule for multi-component HEAs because of the absence of information about multi-component alloys in phase diagrams [3].

It has been argued that alloys with a higher number of principal elements will more easily yield the formation of random solid solutions during solidification, rather than intermetallic compounds, except for those with very large heats of formation [13].

For the alloy system, the Gibbs free energy of mixing can be expressed as follows:

$$\Delta G_{mix} = \Delta H_{mix} - T\Delta S_{mix} \quad (1)$$

Where $\Delta G_{mix}$ is the Gibbs free energy of mixing, $\Delta H_{mix}$ and $\Delta S_{mix}$ are the enthalpy and entropy of mixing, respectively, and *T* is the absolute temperature. From equation (1), it can be seen that if $\Delta H_{mix}$ is kept constant, a higher entropy of mixing will lead to a lower Gibbs free energy, and make the alloy system more stable [14]. For equiatomic systems, the entropy of mixing decreases from -13.381 $JK^{-1}mol^{-1}$ to -17.288 $JK^{-1}mol^{-1}$, as the number of elements increases from 5 to 8. In this paper it is analyzed if the percentage of systems that would form solid solution increases with the number of elements, and the type of solid solutions that would be formed. The following procedure was used.

Multi-component HEAs may form solid solution phases and/or intermetallic compounds or amorphous phases. A solid-solution formation rule for multi-component alloys has been proposed [3]. $\Omega \geq 1.1$ and $\delta \leq 6.6\%$ should be expected as the criteria for forming high entropy stabilized solid solutions. $\Omega$ was defined as a parameter of the entropy of mixing timing the average melting temperature of the elements over the enthalpy of mixing. $\delta$ was defined as the mean square deviation of the atomic size of elements. On the other hand, an analysis of the relationship between valence electron concentration, VEC, and the stability of fcc and bcc phases of the HEAs $Al_xCoCrCuFeNi$, $Al_xCrCuFeNi_2$ and $Al_xCoCrCu_{0.5}FeNi$ was performed. It was found that the stability of fcc and bcc solid solutions can be well delineated by valence electron concentration, VEC. VEC is the number of total electrons accommodated in the valence band, including the *d*-electrons [15]. It was found that at VEC ≥ 8.0 only fcc phase exists, at 6.87 ≤ VEC < 8.0 fcc and bcc were found to coexist, and sole bcc phase exists at VEC < 6.87.

Using the formation rule and the observation stated in the above paragraph, the possibility of solid solution formation in HEAs was analyzed in this work. Thirty elements were included: transition elements of the fourth, fifth and sixth periods of the periodic table, plus aluminum. With these 30 elements, 27,405 alloys with four, 142,506 HEAs with five, 593,775 HEAs with six, and 2,035,800 HEAs with seven elements were formed. Each of these numbers corresponds to the possible combination of 30 elements in sets of four, five, six and seven elements. Values of $\Omega$, $\delta$ and VEC for HEAs with five, six and seven elements from the thirty previously mentioned were calculated. Values of these parameters were also calculated for alloys with four elements in order to compare the results with those of HEAs with five to seven elements. This procedure yielded the number of HEAs that could have fcc, bcc, and a combination of fcc and bcc solid solution phases. In a second step of this analysis, values of $\Omega$, $\delta$ and VEC for alloys with four to seven elements, from the thirty previously mentioned, were also calculated, but maintaining one constant element. With four, five, six and seven elements 3,654, 23,751, 118,755 and 475,020 alloys can be formed in this way, respectively. Alloys having $\Omega \geq 1.1$ and $\delta \leq 6.6\%$ were separated and ranked according to their VEC values. For each of the thirty elements, the percentages of solid solutions with VEC ≥ 8.0, 6.87 ≤ VEC < 8.0 and VEC< 6.87 were calculated. Results are presented as plots of these percentages vs VEC of each constant element.

## 2. Theory.

The enthalpy of mixing for a multi-component alloy system with *n* elements was calculated using the following equation [16]:

$$\Delta H_{mix} = \sum_{i=1, i\neq j} \Omega_{ij} c_i c_j \tag{2}$$

where $\Omega_{ij} = 4\,\Delta \mathrm{H}_{mix}^{AB}$ is the regular solution interaction parameter between the *i*th and *j*th elements, $c_i$ or $c_j$ is the atomic percentage of the **i**th or *j*th component, and $\Delta \mathrm{H}_{mix}^{AB}$ is the enthalpy of mixing of the binary liquid alloys. The values of enthalpy of mixing were obtained from ref. [17], and they are based on the Miedema macroscopic atom model.

The entropy of mixing of an n-element regular solution was calculated as follows

$$\Delta S_{mix} = -R \sum_{i=1}^{n} (c_i ln c_i) \tag{3}$$

where $c_j$ is the atomic percentage of the *i*th component, $\sum_{i=1}^{n} c_i = 1$, and $R$ = 8.314 $JK^{-1}mol^{-1}$ is the gas constant.

For predicting the solid-solution formation for various multi-component alloys, a parameter Ω was defined as:

$$\Omega = \frac{T_m \Delta S_{mix}}{|\Delta H_{mix}|} \tag{4}$$

The melting temperature of n-element alloy, $T_m$, was calculated using the rule of mixtures:

$$T_m = \sum_{i=1}^{n} c_i (T_m)_i \tag{5}$$

where $(T_m)_i$ is the melting point of the *i*th component of alloy. $T_m$ values were taken from reference [18].

A parameter$\delta$, which describes the compressive effect of the atomic size difference in n-element alloy, was calculated as:

$$\delta = 100\sqrt{\sum_{i=1}^{n} c_i \left(1 - \frac{r_i}{\bar{r}}\right)^2} \tag{6}$$

where $c_j$ is the atomic percentage of the *i*th component, $\bar{r} = \sum_{i=1}^{n} c_i r_i$ is the average atomic radius and $r_i$, the atomic radius, was obtained from reference [19].

The effect of valence electron concentration on the stability of fcc and bcc solid solutions formed in HEAs has been analyzed [15]. It was found that fcc phases are stable at higher VEC (≥ 8.0), and, instead, bcc phases are stable at lower VEC (<6.87). In the range 6.87 ≤ VEC < 8.0 mixed fcc and bcc phases will exist.
For HEAs, VEC was defined as [15]:

$$VEC = \sum_{i=1}^{n} c_i (VEC)_i \tag{7}$$

where $(VEC)_i$ is VEC for the individual element. VEC values were taken from reference [20].

## 3. Results.

Table 1 presents the total number of alloys that can be formed with four to seven elements and the percentage of alloys that fulfill the rule Ω ≥ 1.1 and δ ≤ 6.6. It can be seen that the proportion of alloys that would form solid solutions decreases from 35.87% with four elements to 26.42% with seven elements. As the number of elements increases, the percentage of alloys with a combination of bcc and fcc phases would increase slightly from 9.28 to 11.06. Conversely, the proportion of alloys with fcc or bcc structures would diminish as the number of elements increases. From equation (1), it could be expected that the decrease in enthalpy of mixing would increase the percentage of alloys with solid solution structure. This, as the number of elements in the alloys increases. However, the results in table 1 show that the opposite behavior could be expected.

Table 1. Number of alloys that can be formed, and percentage of these that would form solid solutions.

| Number of elements | Total number of alloys that can be formed | Alloys with Ω ≥ 1.1 and $\delta$ ≤ 6.6 (%) | Alloys with Ω ≥ 1.1 $\delta$ ≤ 6.6 (%) and: VEC (<6.87) | 6.87 ≤ VEC < 8.0 | VEC (≥ 8.0) |
|---|---|---|---|---|---|
| 4 | 27,405 | 9,724 (35.87) | 7.03 | 9.28 | 19.16 |
| 5 | 142,506 | 44,713 (31.37) | 5.55 | 10.18 | 15.64 |
| 6 | 593,775 | 169,265 (28.51) | 4.62 | 10.05 | 13.84 |
| 7 | 2,035,800 | 537,828 (26.42) | 3.08 | 11.06 | 12.28 |

The ability of a given element to form alloys with a solid solution structure was evaluated as follows. The percentage of alloys having Ω ≥ 1.1 and δ ≤ 6.6 was calculated for alloys with four to seven elements, maintaining one constant element. For alloys with seven elements, the plot of percentage of alloys with solid solution structure vs VEC of the element maintained constant is shown in figure 1. It can be seen that there is a weak linear correlation between the percentage and VEC of the element, $R^2$= 0.103. Alloys with four, five and six elements presented also low correlation coefficients, $R^2$=0.146, 0.102and 0.104, respectively. Figure 2 shows the plot of percentage of alloys with Ω ≥ 1.1 and δ ≤ 6.6 vs the absolute value of the difference of $r$ minus $r_{av}$, for alloys with seven elements. $r$ is theradius of the element maintained constant, and $r_{av}$ is the average radius of the thirty elements. A linear correlation coefficient, $R^2$= 0.781, was calculated.The calculated correlation coefficients for alloys with four, five and six elements were 0.594, 0.678 and 0.762, respectively.

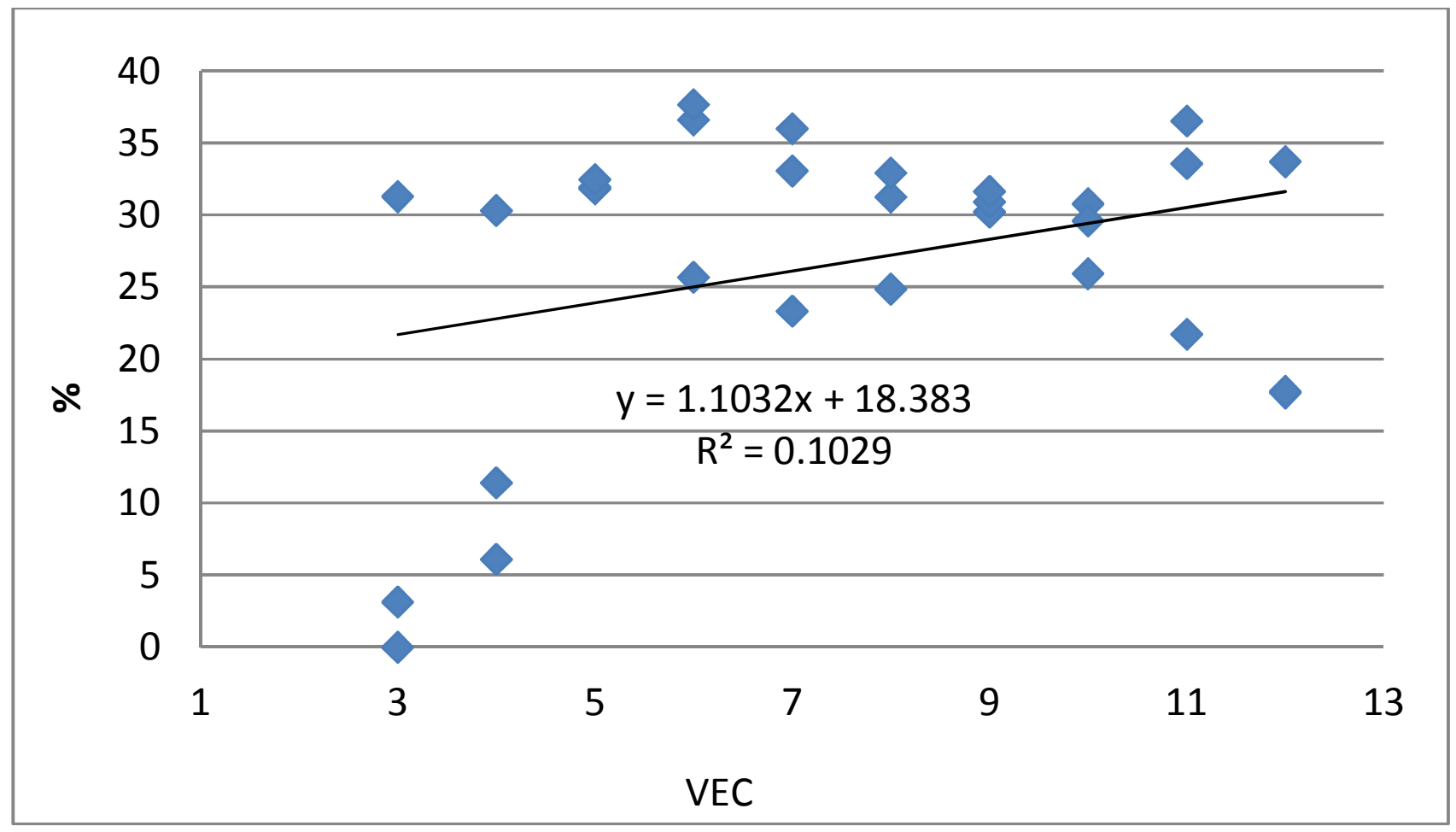

Figure 1. Plot of % of alloys with Ω ≥ 1.1 and δ ≤ 6.6 vs. VEC of the element maintained constant, for alloys with 7 elements.

It can be observed that the ability of a given element to form alloys with a solid solution structure decreases as the radius of the element increases. In fact, Y, the biggest of the thirty elements considered, would form a very small number of solid solutions: 6, 2, 14 and 9 for alloys with 4, 5, 6 and 7 elements, respectively. 31 alloys containing Y with solid solution structure, out of the total of 621,180 alloys that would be formed with this element.

The effect of atomic size on the formation of alloys that would have bcc, fcc or a combination of bcc and fcc solid solutions was also evaluated. In figure 2, it can be observed that there is a linear relationship between the percentage of alloys that would form solid solution and the absolute value of the difference between the radius of the element maintained constant minus average radius.However, when the percentage of alloys that would form solid solutions is decomposed according to the expected type of solid solution, this same relationship is weak. Figure 3 shows the plot corresponding to alloys containing 7 elements. Weak linear correlation coefficients can be observed in the three cases. $R^2$= 0.361, 0.256 and 0.195 for alloys with VEC < 6.87, 6.87 ≤ VEC < 8.0, and VEC > 8.0, respectively. The plots corresponding to alloys with 6, 5 and 4 elements are omitted because they are similar to the one with 7 elements.

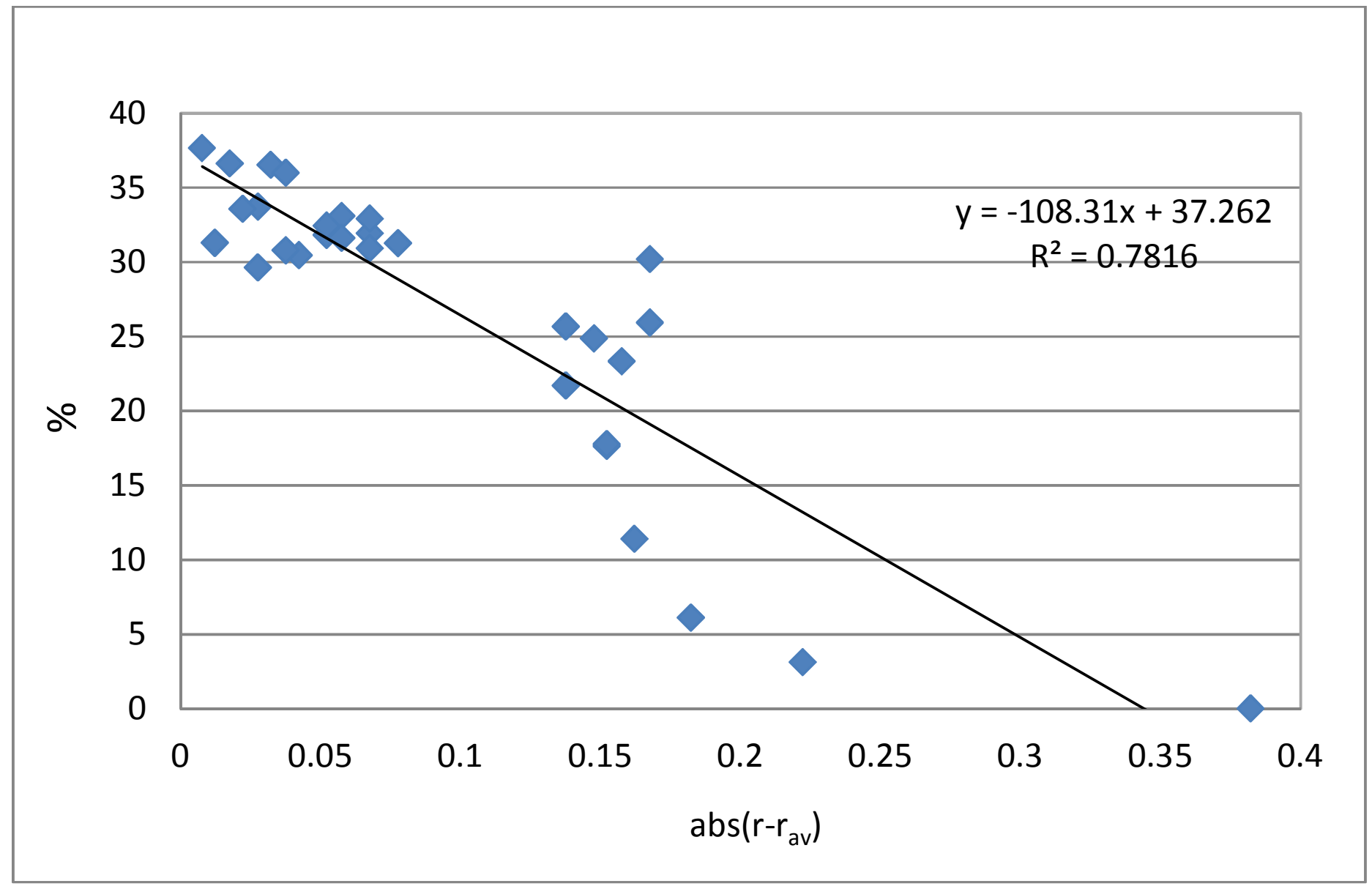

Figure 2. Plot of % of alloys with Ω ≥ 1.1 and δ ≤ 6.6 vs. the absolute value of the difference between the radius of the element maintained constant minus average radius, for alloys with 7 elements.

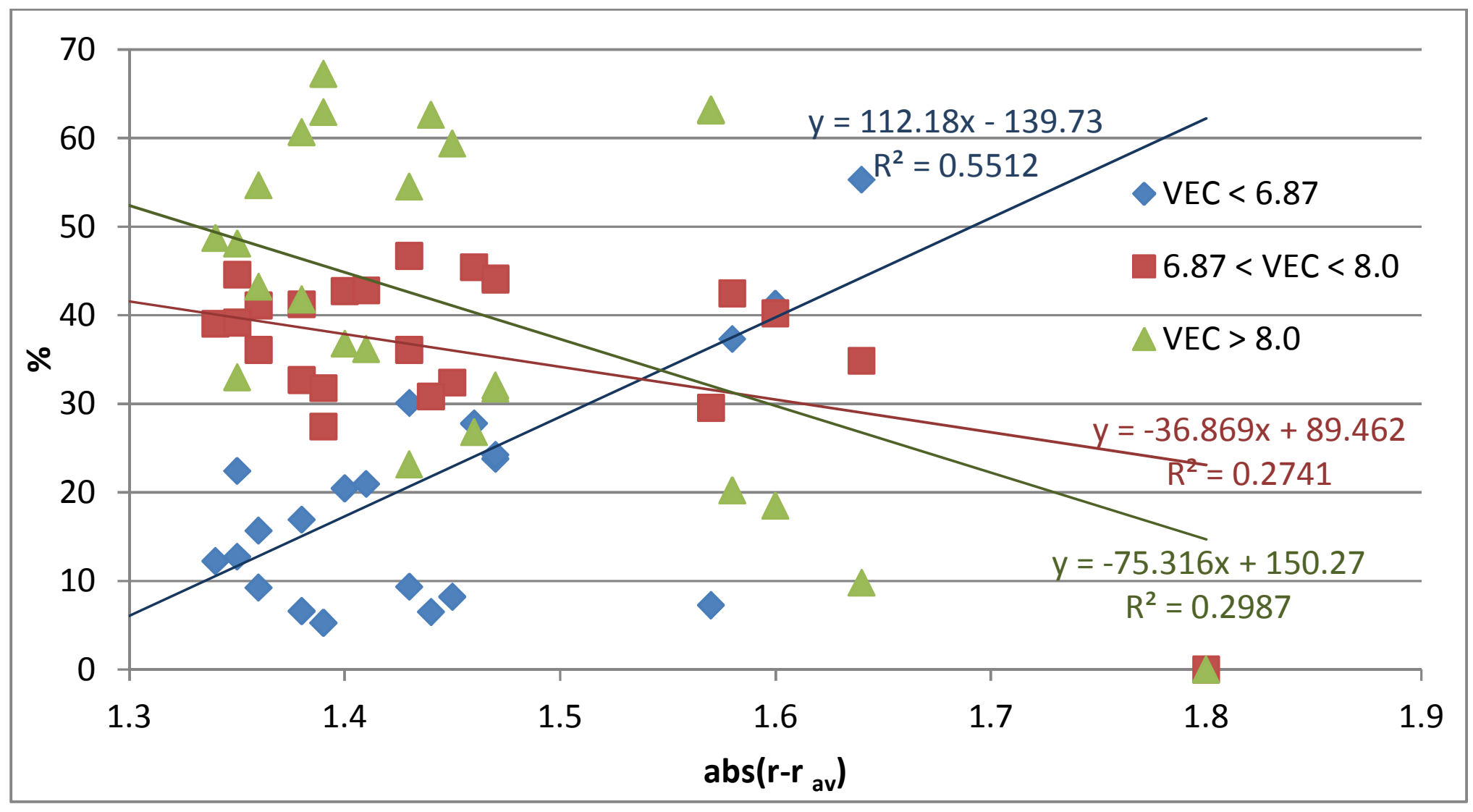

Figure 3 Plot of percentage of alloys that would form solid solutions vs. the absolute value of the difference between the radius of the element maintained constant minus average radius, for alloys with 7 elements.

Above, it was stated that there is a weak linear relationship between the percentage of alloys that would form solid solutions and VEC of the element maintained constant. However, when this percentage is decomposed according to the type of solid solution, the relationship between these two variables can be expressed with a quadratic model. The agreement with this model is better for alloys that would have a bcc or fcc solid solution than for alloys with a combination of bcc and fcc solid solutions. Figure 4 shows the plot for alloys with 7 elements. $R^2$= 0.956, 0.713 and 0.336 for VEC > 8.0, < 6.87 and 6.87 ≤ VEC < 8.0, respectively. The correlation coefficient for bcc solid solutions, compared with the coefficient for fcc solid solutions, is lower due to Y. This element would form only bcc solid solutions because of its combination of atomic radius and VEC.

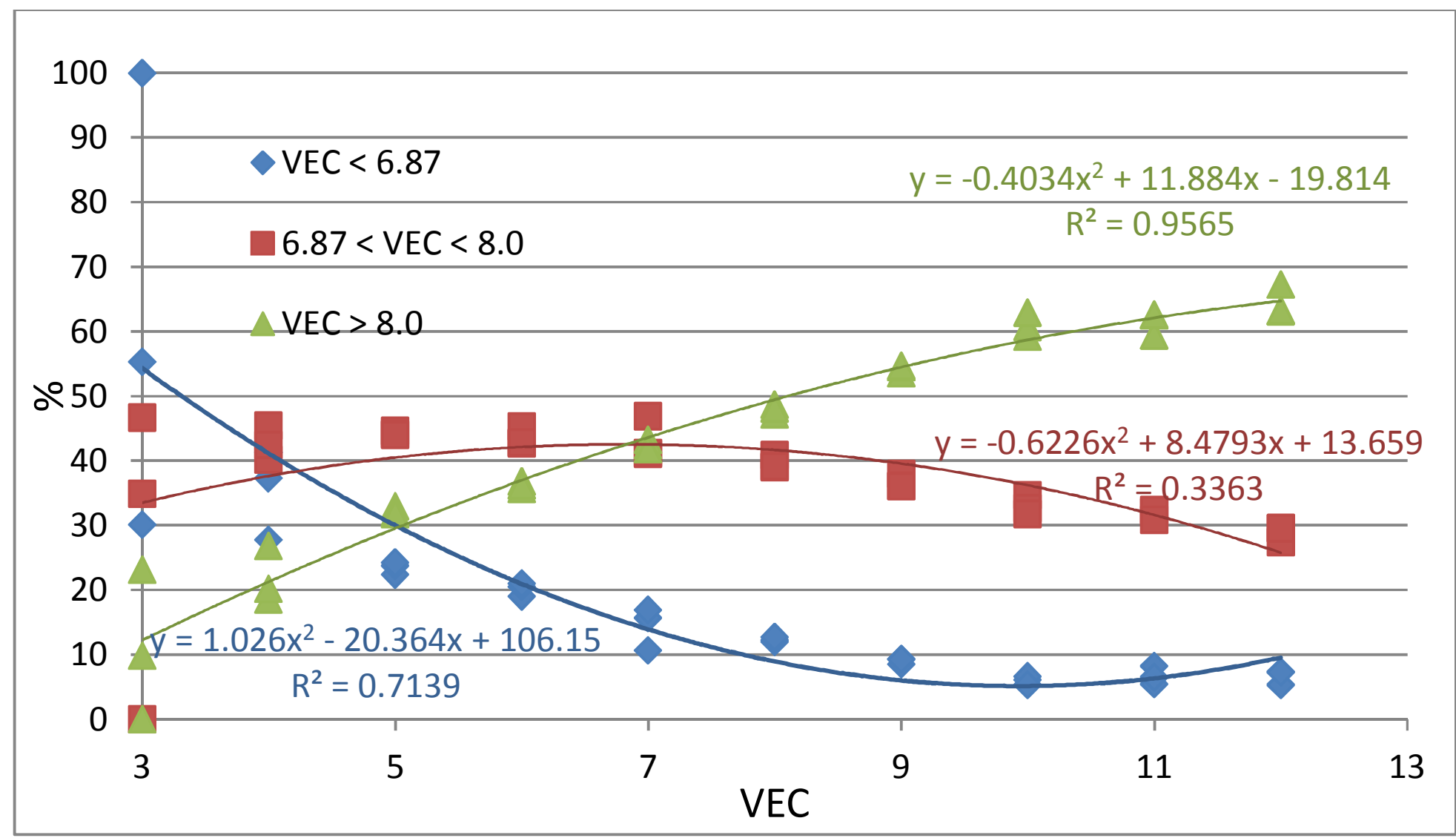

Figure 4. Plot of percentage of alloys that would form solid solutions vs. VEC, for alloys with VEC < 6.87, 6.87 ≤ VEC < 8.0, and VEC > 8.0, corresponding to alloys with 7 elements.

Figure 5 shows the variation of percentage of alloys that would form solid solutions vs. VEC of the element maintained constant, for alloys with VEC < 6.87. It can be observed that for alloys of elements with VEC equal to 3 or 4, there is a high dispersion in the percentage of alloys that would form bcc solid solutions. For elements with VEC of 3, the percentage fluctuates from around 30 to 100 %. For elements with VEC from 8 to 12 the variation on this percentage is small. The proportion of alloys that would have a bcc structure decreases with the increase in VEC, for elements with VEC from 3 to 10. Then, this proportion increases slightly for alloys with VEC of 11 and 12. It is interesting to note that alloys containing elements with VEC from 9 to 12 would form reduced proportions of alloys with bcc structure, around 10%.

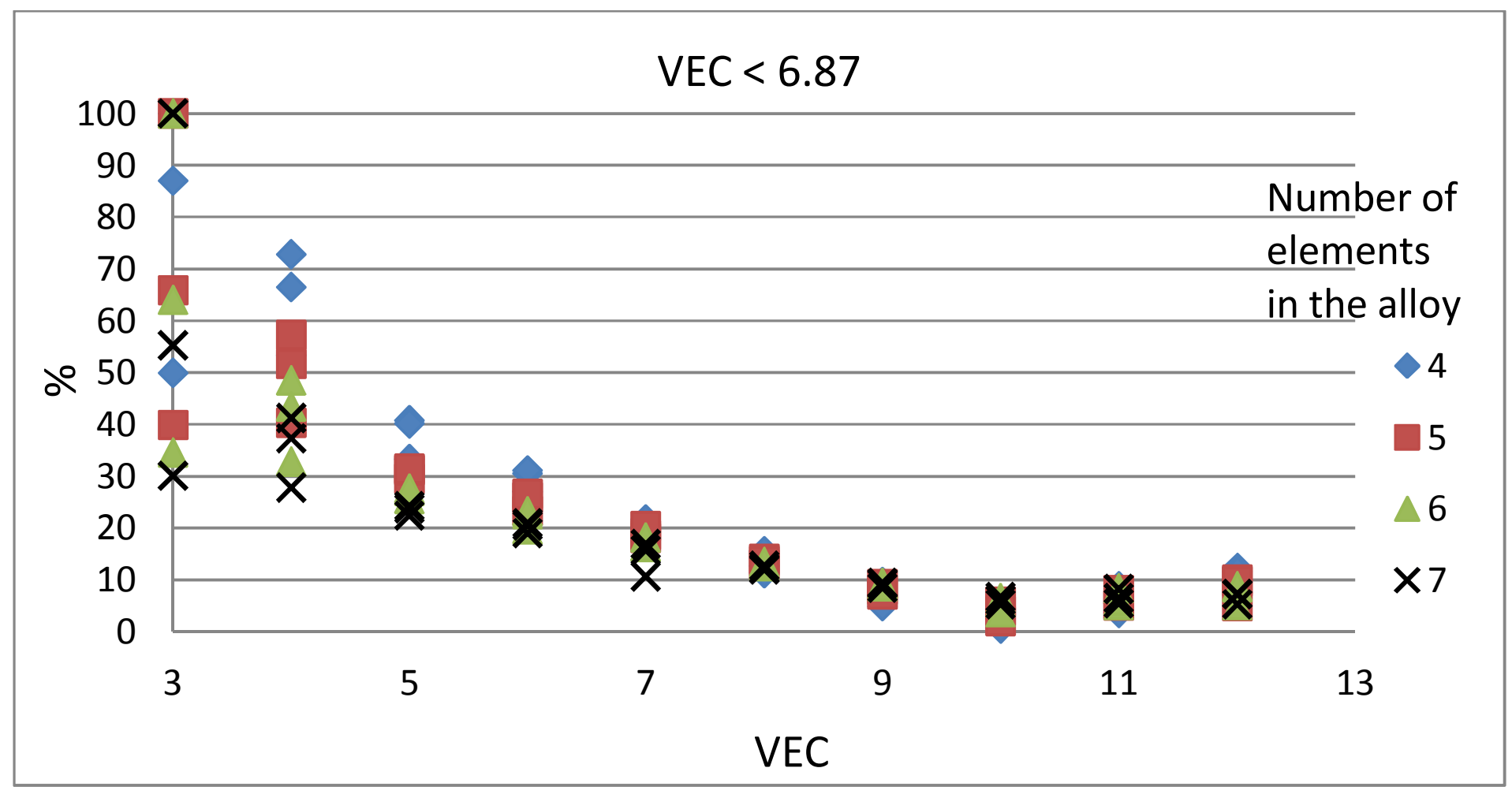

Figure 5. Percentage of alloys that would form solid solutions vs. VEC, for alloys with VEC < 6.87, corresponding to alloys with 4 to 7 elements.

In figure 6, the percentage of alloys that would form solid solutions vs. VEC, for alloys with 6.87 ≤ VEC < 8.0 is plotted. It can be observed that alloys with VEC of 3 or with 4 elements show a high dispersion of the percentage of alloys that would have bcc and fcc structures. In general terms, the proportion of alloys that would have a mixture of fcc and bcc solid solutions increases as VEC does, for alloys with VEC from 3 to 7. Then, this proportion diminishes as VEC increases from 8 to 12.

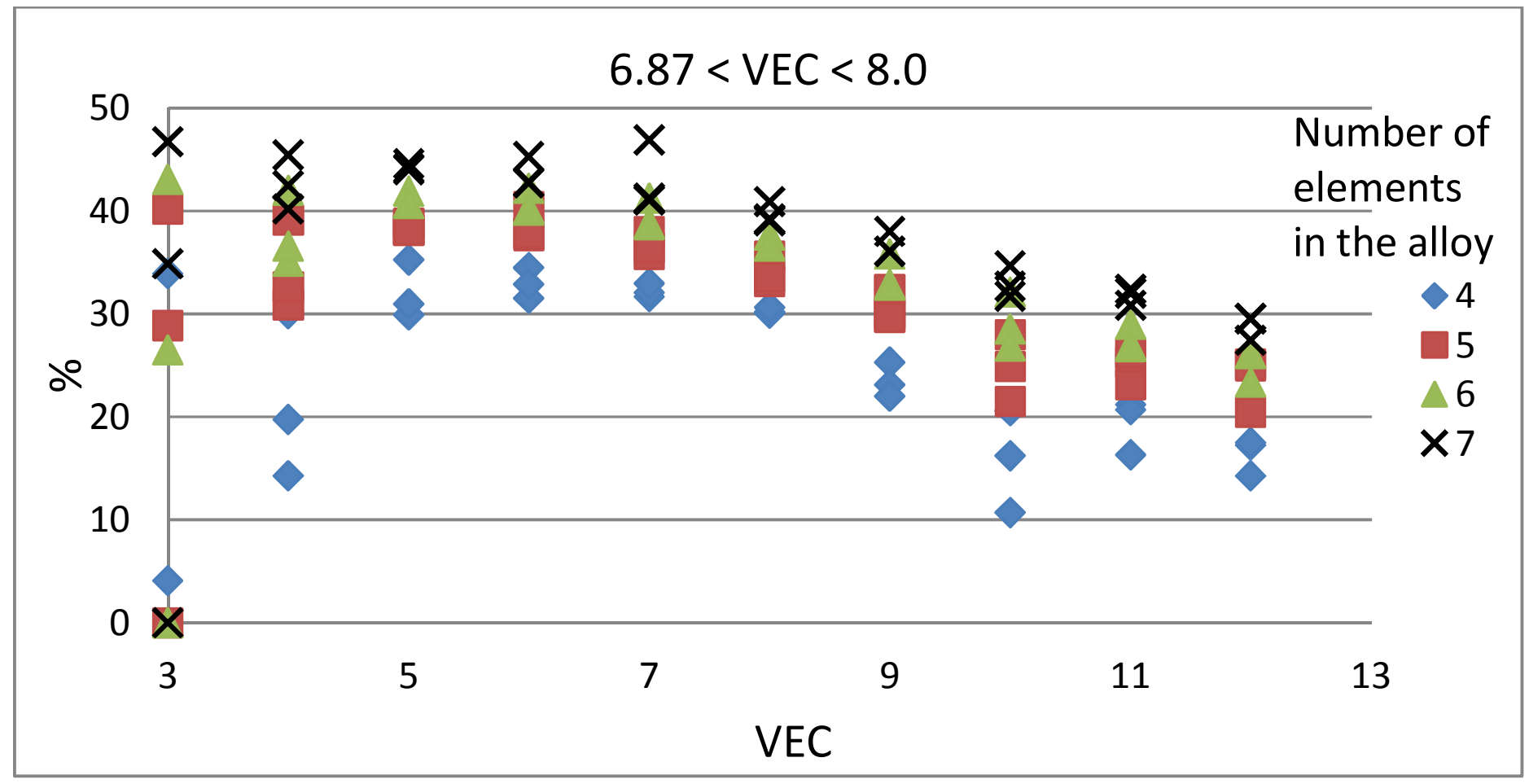

Figure 6. Percentage of alloys that would form solid solutions vs. VEC, for alloys with 6.87 ≤ VEC < 8.0, corresponding to alloys with 4 to 7 elements.

Figure 7 shows that the percentage of alloys that would form fcc solid solutions, VEC > 8.0, increases with VEC, for elements with VEC from 3 to 10. This percentage increases from an average value of around 10% for alloys with VEC of 3, to an average of around 75% for alloys with VEC of 10. For alloys with VEC of 11 and 12, the percentage diminishes a little to an average of around 70%. Zn, Pd, Pt and Au would form a high percentage of alloys with four elements and fcc structure, 80.6, 81.3, 88.7 and 80.0%, respectively.

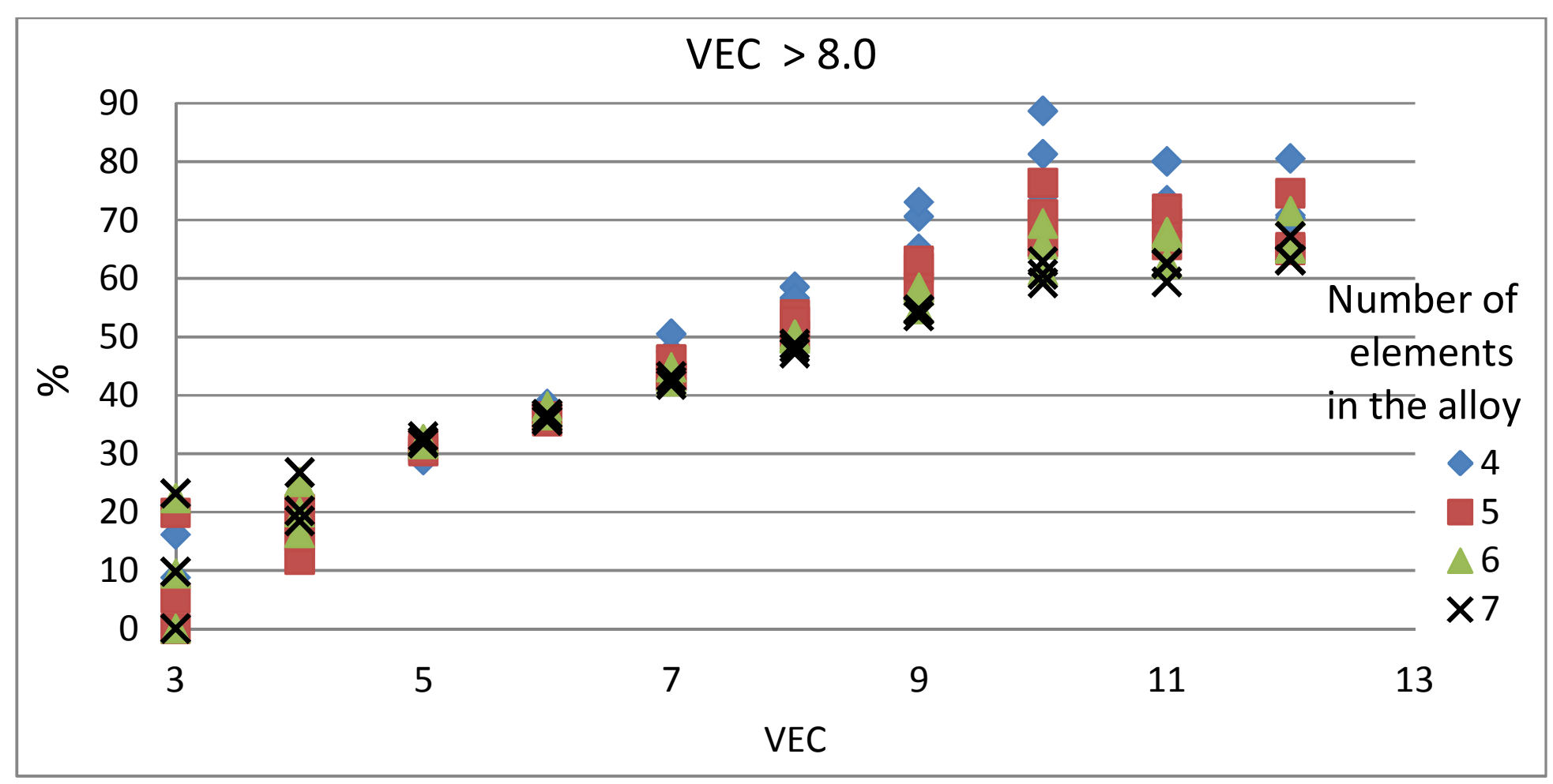

Figure 7. Percentage of alloys that would form solid solutions vs. VEC, for alloys with VEC > 8.0, corresponding to alloys with 4 to 7 elements.

Figure 5 shows that elements with VEC of 3 and 4 would have a high percentage of solid solutions with bcc structure. On the other hand, elements with VEC of 9 to 12 would have only around 10% of alloys with bcc structure. In figure 7, it is clear that the opposite behavior would be observed in the formation of alloys with fcc structure. The effect of valence electron per atom (e-/a), on the stability of fcc and bcc transition elements and binary intermetallic compounds has been analyzed [21]. It was concluded that for transition elements, with 4 or 5 d-electrons per atom the bcc structure is stable, while with 6 to 9 d-electrons per atom the fcc structure is stable. Then, maybe elements with lower VECs would form higher percentages of alloys with bcc structure because their alloys would have lower VECs also. Elements with higher VECs would form higher percentages of alloys with fcc structure because their alloys would have higher VECs also. The role of VEC on the structure of binary solid solutions can be illustrated with table 2. In this table, VECs of twenty two fcc and seven bcc binary solid solutions, arbitrarily selected, are presented. Nineteen binary solid solutions with fcc structure have VEC ≥ 8.0, and three present VEC < 8.0. Six solid solutions with bcc structure show VEC < 6.87. Table 2 also shows VEC values for nine intermetallic compounds with CsCl-type structure, and for nine intermetallic compounds with $Cu_3Au$-type structure. These intermetallic structures are derived from the bcc and the fcc structures, respectively. Six out of the nine intermetallics with $Cu_3Au$ structure have VEC ≥ 8.0, and three have VEC < 8.0. On the other side, three out of the nine intermetallics with CsCl structure have VEC ≥ 8.0, and six have VEC < 6.87.

In table 2, values of Ω and δ for the above mentioned solid solutions and intermetallic compounds are also reported. For four bcc and four fcc solid solutions, Ω could not be calculated because $\Delta H_{mix} = 0$ in these systems. Eight out of the eighteen fcc solid solutions fulfill the rule of Ω ≥ 1.1 and δ ≤ 6.6 for solid solution formation in HEAs. Three bcc systems have Ω and δ values, and they fulfill the rule. Only one out of eighteen intermetallic compounds has Ω ≥ 1.1 and δ ≤ 6.6. Although the rule was not derived for binary solid solutions or intermetallic compounds, it seems that it could be applied to binary intermetallic compounds. The number of binary solid solutions with Ω and δ values in the table is too small to draw any conclusion.

Table 2. Lattice type, VEC, $\delta$ and Ω for binary solid and intermetallic compounds arbitrarily selected.

| System | Lattice | VEC | $\delta$ | Ω |
|---|---|---|---|---|
| $Co_{0.5}Pt_{0.5}$ | fcc | 9.5 | 5.30 | 1.57 |
| $Ni_{0.5}Co_{0.5}$ | fcc | 9.5 | 0.00 | |
| $Co_{0.7}Pt_{0.3}$ | fcc | 9.3 | 4.97 | 1.60 |
| $Mn_{0.3}Pd_{0.7}$ | fcc | 9.1 | 4.09 | 0.46 |
| $AgMg_{0.5}Zn_{0.5}$ | fcc | 9 | 5.30 | |
| $Co_{0.92}Fe_{0.08}$ | fcc | 8.92 | 0.43 | 13.95 |
| $Co_{0.8}Os_{0.2}$ | fcc | 8.8 | 3.15 | |
| $Co_{0.8}Ru_{0.2}$ | fcc | 8.8 | 2.84 | 12.58 |
| $Mg_{0.25}Ag_{0.75}$ | fcc | 8.75 | 4.39 | 0.72 |
| $Pt_{0.36}Fe_{0.64}$ | fcc | 8.7 | 4.39 | 0.86 |
| $Co_{0.9}Cr_{0.1}$ | fcc | 8.7 | 0.72 | 3.40 |
| $Co_{0.8}Re_{0.2}$ | fcc | 8.6 | 4.08 | 6.85 |
| $Cu_2FeSn$ | fcc | 8.5 | 10.92 | 0.85 |
| $Co_2GeZn$ | fcc | 8.5 | 4.85 | 0.69 |
| $Mn_{0.5}Pt_{0.5}$ | fcc | 8.5 | 4.91 | 0.37 |
| $Co_{0.8}W_{0.2}$ | fcc | 8.4 | 4.99 | 14.00 |
| $Rh_{0.7}Mn_{0.3}$ | fcc | 8.4 | 3.12 | 0.76 |
| CoRe | fcc | 8 | 4.94 | 7.53 |
| $Co_{0.75}V_{0.25}$ | fcc | 8 | 3.40 | 0.83 |
| $Co_{0.8}Ga_{0.2}$ | fcc | 7.8 | 4.44 | 0.87 |
| $Mn_{0.5}Fe_{0.5}$ | fcc | 7.5 | 0.40 | |
| $Li_{0.4}Au_{0.6}$ | fcc | 7 | 2.63 | 0.15 |
| $MnPd_3$ | $Cu_3Au$ -type | 9.25 | 3.85 | 0.47 |
| $MgAg_3$ | $Cu_3Au$ -type | 8.75 | 4.39 | 0.72 |
| $Au_3Li$ | $Cu_3Au$ -type | 8.5 | 2.34 | 0.19 |
| $Ni_3Ga$ | $Cu_3Au$ -type | 8.25 | 4.78 | 0.57 |
| $CrRh_3$ | $Cu_3Au$ -type | 8.25 | 2.27 | 1.07 |
| $TaRh_3$ | $Cu_3Au$ -type | 8 | 3.77 | 0.35 |
| $Mn_3Pt$ | $Cu_3Au$ -type | 7.75 | 4.36 | 0.37 |
| $Mn_3Rh$ | $Cu_3Au$ -type | 7.5 | 3.04 | 0.66 |
| $Co_{0.7}V_{3.3}$ | $Cu_3Au$ -type | 5.7 | 2.85 | 1.01 |
| $Cr_{0.7}Co_{0.3}$ | bcc | 6.9 | 1.08 | 3.11 |
| $Cr_{0.8}Ni_{0.2}$ | bcc | 6.8 | 0.94 | 1.94 |
| $La_{0.8}Mg_{0.2}$ | bcc | 2.8 | 6.10 | 1.06 |
| $Mg_{0.7}Li_{0.3}$ | bcc | 1.7 | 2.38 | |
| $Li_{0.3}Mg_{0.7}$ | bcc | 1.7 | 2.38 | |
| $Li_{0.5}Mg_{0.5}$ | bcc | 1.5 | 2.63 | |
| $Li_{0.9}Mg_{0.1}$ | bcc | 1.1 | 1.61 | |
| MnPd | CsCl -type | 8.5 | 4.55 | 0.42 |
| CoFe | CsCl -type | 8.5 | 0.79 | 10.31 |
| MnRh | CsCl -type | 8 | 3.45 | 0.68 |
| MgAg | CsCl -type | 6.5 | 4.95 | 0.62 |
| MgAu | CsCl -type | 6.5 | 5.29 | 0.20 |
| NiAl | CsCl -type | 6.5 | 6.72 | 0.35 |
| CoGa | CsCl -type | 6 | 5.37 | 0.54 |
| LiAu | CsCl -type | 6 | 2.67 | 0.14 |
| MgLa | CsCl -type | 2.5 | 7.99 | 0.87 |

**Conclusions.**

The possibility of solid solution formation in high entropy alloys (HEAs) has been calculated for alloys with four to seven elements. A rule previously reported, $\Omega \geq 1.1$ and $\delta \leq 6.6\%$, was used as the criteria for forming high entropy stabilized solid solutions. Thirty elements were included: transition elements of the fourth, fifth and sixth periods of the periodic table, and aluminum. A total of 2,799,486 systems were analyzed. The percentage of solid solutions that would be formed in HEAs decreases from 35.9% to 26.4%, as the number of elements increases from four to seven. The structure of the solid solutions, fcc, bcc or a mixture of fcc and bcc, that would be formed, has been predicted using a previously reported observation. According to this observation, fcc phases are stable at higher VEC (≥ 8.0), and bcc phases are stable at lower VEC (<6.87). In the range 6.87 ≤ VEC < 8.0 mixed fcc and bcc phases will exist. The percentage of systems with fcc or bcc structure decreases as the number of elements increases from four to seven. The percentages of solid solutions with fcc, bcc or a mixture of fcc and bcc was calculated, for alloys with four to seven elements, but maintaining one constant element. Systems with VEC from three to four, would have bcc structure in around 50% of the systems. In systems with VEC from ten to twelve, around 75% of systems would present fcc structure.